\documentclass[seceq,epsfig]{ptptex}
\usepackage{epsfig}

\title{
Four- and Five-Body Scattering Calculations of \\
Exotic Hadron Systems%
}

\author{
Emiko \textsc{Hiyama,}$^{1}$\footnote{ e-mail address:
hiyama@cc.nara-wu.ac.jp} 
Hideo \textsc{Suganuma}$^{2}$
and  
Masayasu \textsc{Kamimura}$^{3}$
}

\inst{
$^1$ Department of Physics,
Nara Women's University, Nara 630-8506, Japan\\
$^{2}$ Department of Physics, Kyoto University,
Kyoto, 606-8502, Japan\\
$^{3}$ Department of Physics, Kyushu University,
Fukuoka, 812-8581, Japan
}

\abst{
We study the five-quark system $uudd{\bar s}$  
in the standard non-relativistic quark model 
by solving the scattering problem.
Using the Gaussian Expansion Method (GEM), 
we perform the almost precise multi-quark calculations 
by treating a very large five-body modelspace 
including the NK scattering channel explicitly. 
Although a lot of pseudostates (discretized continuum states) 
with $J^\pi=\frac{1}{2}^\pm$ and $J^\pi=\frac{3}{2}^\pm$  
are obtained within the bound-state approximation, 
all the states in $1.4-1.85$ GeV in mass around ${\rm {\rm \Theta}}^+(1540)$ 
melt into non-resonant continuum states 
through the coupling with the NK scattering state
in the realistic case, i.e., 
there is no five-quark resonance below 1.85GeV.
Instead, we predict a five-quark resonance state of $J^\pi=\frac{1}{2}^-$ 
with the mass of about 1.9GeV and the width of $\Gamma \simeq$ 2.68MeV.
Similar calculation is done for the four-quark system $c{\bar c}q{\bar q}$ ($q=u,d$)
in connection with X$^0$(3872). 
}

\begin{document}

\maketitle

\section{Introduction: importance of the wave function of multi-quarks}

From the viewpoint of Elementary Particle Physics, 
the lattice QCD calculation is one of the most standard approaches 
to investigate hadrons including multi-quarks. 
However, it is rather difficult to extract the ``wave function" of hadrons 
in the lattice QCD calculation because it is based on the path-integral, 
where all contributions of possible states are summed up and 
only vacuum expectation values can be obtained.

Needless to say, the wave function is one of the most important quantities 
in quantum physics, and, of course, also in Quark-Hadron Physics.
In particular for multi-quark systems, 
the analysis with the quark wave function is necessary to clarify whether 
the multi-quark hadron is an exotic resonance state or a two-hadron scattering state.
Thus, to extract the state information (the wave function) of hadrons, 
we need a reliable calculational method for multi-quark systems instead of lattice QCD.
One of the attractive methods is the constituent quark-model calculation,  
since it seems workable up to several hundred MeV excitation, 
and its precise calculation for multi-quark systems can be done with 
the Gaussian Expansion Method (GEM) \cite{Hiyama2003,Hiyama2006}.

In Nuclear Physics, 
a precise calculational method for bound and scattering states
of various few-body systems using GEM \cite{Hiyama2003}
has been developed by two of the present
authors (E.H. and M.K.) and their collaborators,
and has successfully been applied to light nuclei, light hypernuclei,
exotic atoms/molecules and so on \cite{Hiyama2003}.

Using GEM within the framework of a constituent quark model, 
Hiyama {\it et al.}\cite{Hiyama2006} investigated 
for the first time scattering and resonance states 
of the five-quark system $uudd{\bar s}$ 
under the explicit NK scattering boundary condition.
It was made clear that there appears no $\frac{1}{2}^\pm$ resonance state
around the reported energy of ${\rm \Theta}^+(1540)$ 
\cite{Nakano03,Dzierba05,Oka04}.

In this paper, we perform the similar precise calculation of 
the five-quark system $uudd{\bar s}$ with $J^\pi=\frac{1}{2}^\pm$ and $J^\pi=\frac{3}{2}^\pm$ 
using the linear confinement potential, 
which is more appropriate as indicated by lattice QCD calculations \cite{TS0102,Okiharu}.
Furthermore, we apply the same method to the four-quark 
system $c{\bar c}q{\bar q}$ in connection with
X$^0$(3872) \cite{X3872} and  
discuss about  possibility of any bound/resonance 
state near the ${\rm D}^0 {\rm \bar D}^{*0}$ threshold. 

\section{Five-body quark-model calculation for ${\rm \Theta}^+(1540)$}

We study the five-quark system $uudd{\bar s}$ 
by solving the five-body Schr\"{o}dinger equation
$
( H - E )\, \Psi_{J^\pi M}  = 0     
$
including the NK scattering channel explicitly, 
as was done in Ref.2).
We here take a standard non-relativistic quark-model Hamiltonian
with the linear-type confining potential $V_{\rm conf}$ and
the color-magnetic potential $V_{\rm CM}$: 
\begin{equation}
V_{\rm conf}(r_{ij}) = - 
 \frac{\lambda^a_i}{2}\,\frac{\lambda^a_j}{2}
 (\frac{3}{4}\sigma~r_{ij} + v_0 ), \  
V_{\rm CM}(r_{ij})=-
 \frac{\lambda^a_i}{2}\,\frac{\lambda^a_j}{2}
 {{\xi_{\alpha}} \over {m_i m_j}}\,
e^{- r_{ij}^2/ \beta^2 }
\, \mbox{\boldmath $\sigma$}_i \cdot
  \mbox{\boldmath $\sigma$}_j.
\label{Vcm}
\end{equation}
Using the parameters listed in Table I, 
we can well reproduce masses of ordinary baryons and mesons as shown in Table II and III, 
and well reproduce or predict hadron properties, {\it e.g.},
$\mu_{\rm p} \simeq  2.75{\rm nm}$ (exp. 2.78nm),
$\mu_{\rm n} \simeq -1.80{\rm nm}$ (exp. $-$1.91nm),
$\mu_{\rm \Lambda} \simeq -0.60{\rm nm}$ (exp. $-$0.61nm),
$\mu_{\rm \Sigma^0} \simeq 0.81{\rm nm}$,
$\mu_{\rm \Omega} \simeq  -1.84{\rm nm}$ (exp. $-$2.02nm)
for the magnetic moment.
Note here that the form of the linear-type confining potential $V_{\rm conf}$ is 
appropriate as an approximation of the Y-type linear three-quark potential 
indicated by lattice QCD \cite{TS0102,Okiharu}, 
although the adopted value of the string tension $\sigma \simeq$ 0.56GeV/fm 
is smaller than the standard value $\sigma \simeq$ 0.89GeV/fm.

\begin{table}[h]
\begin{center}
\caption{Parameters of the present constituent quark model with a linear-type confining potential.}
\begin{tabular}{ccccccc}
\hline\hline
parameter     &  $m_u$, $m_d$ & $m_s$  & string tension $\sigma$ & $\beta$ & $v_0$ & $\xi_\alpha$ \\ 
\hline
adopted value &     330MeV    & 500MeV & 0.56 GeV/fm & 0.5fm   & $-$572MeV & 240MeV$\times m_u^2$\\
\hline\hline
\end{tabular}
\end{center}
\vspace{-0.45cm}
\end{table}

\begin{table}[h]
\begin{center}
\caption{Calculated masses of typical baryons obtained with the present constituent quark model.}
\begin{tabular}{cccccccc}
\hline\hline
baryon & N & $\rm \Delta$ & $\rm \Lambda$ & $\rm \Sigma$ & $\rm \Xi$ & $\rm \Omega$ & N* \\ \hline
calculated mass [MeV] & 939 &  1235 & 1064.8& 1129 & 1324 & 1539.5  & 1462  \\
empirical mass [MeV] & 939 & 1232 & 1115.7  & 1191 & 1318 & 1672.5 & 1440  \\
\hline\hline
\end{tabular}
\end{center}
\vspace{-0.45cm}
\end{table}

\begin{table}[h]
\begin{center}
\caption{Calculated masses of typical mesons obtained with the present constituent quark model.}
\begin{tabular}{cccccc}
\hline\hline
meson & $\omega$ & $\rho$ & K* & K & $\pi$\\ \hline
calculated mass  [MeV] & 759 & 759.3 & 864.4 & 457.9 & 152 \\
empirical value  [MeV] & 782.7 & 775.5 & 892 & 496   & 138 \\
\hline\hline
\end{tabular}
\end{center}
\end{table}

\begin{figure}[t]
\centering
\epsfig{file=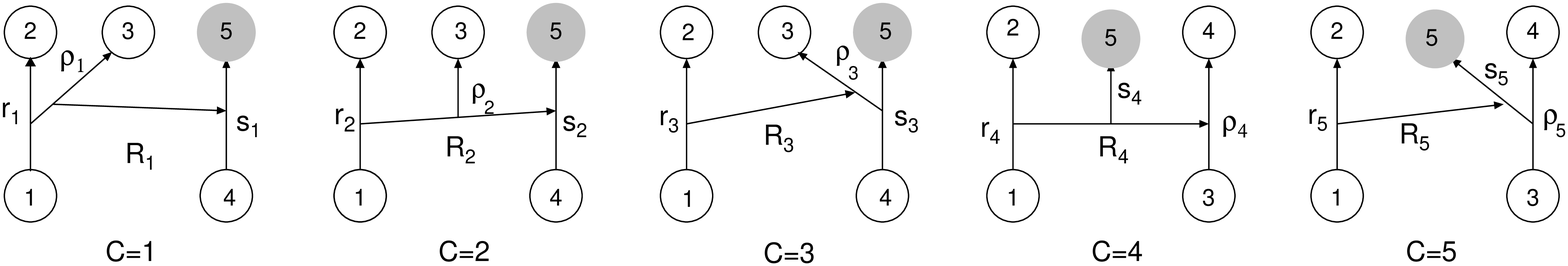,scale=0.27}
\caption{The five important sets of Jacobi coordinates among five quarks. 
Four light quarks ($u, d$), labeled by particle $1-4$, are to be 
antisymmetrized, while particle 5 stands for ${\bar s}$ quark. 
Sets $c=4, 5$ explicitly contain two $qq$ correlations, 
while sets $c = 1-3$ explicitly contain $qq$ and $q{\bar q}$ correlations \cite{Hiyama2006}.  
The NK scattering channel is treated with $c=1$.
} 
\label{fig:pen5-jacobi}
\end{figure}

Now, we perform an almost precise quark-model calculation with GEM 
for multi-quark systems and 
clarify whether each obtained state is a resonance or a continuum scattering state 
through the phase-shift analysis in the model calculation.
For the five-quark system $uudd\bar s$ with the energy $E$, 
the total wave function is given as 
\begin{equation} 
\Psi_{J^\pi M}(E)= \Psi^{\rm (NK)}_{J^\pi M}(E) + 
\sum_{\nu=1}^{\nu_{\rm max}} b_J^{(\nu)}(E)  
\Phi^{(\nu)}_{J^\pi M}(E_\nu).
\end{equation}
The first term is the NK scattering component
expressed by 
\begin{eqnarray}
\Psi_{J^\pi M} ^{\rm (NK)}(E)= {\cal A}_{1234}\!
\big\{  \big[ \big[\phi_{\frac{1}{2}}^{\rm (N)}(123)
 \phi_0^{\rm (K)}(45)\big]_{\frac{1}{2}}   
\chi_L({\bf R}_1)\big]_{J^\pi M} \big\}. \;
\label{elastic}
\end{eqnarray}
The second term describes five-body degrees of freedom in
the interaction-region amplitude which vanishes asymptotically.
The amplitude is expanded using a nearly complete set of
five-body eigenstates,
$\{\Phi^{(\nu)}_{J^\pi M}(E_\nu)\} (\nu=1-\nu_{\rm max})$, 
constructed by diagonalizing the total Hamiltonian as
$
\langle \,\Phi^{(\nu)}_{J^\pi M}(E_\nu) \, | H | \,
\Phi^{(\nu')}_{J^\pi M}(E_\nu') \,\rangle
= E_\nu\, \delta_{\nu \nu'}.
$
Here, each of $\Phi^{(\nu)}_{J^\pi M}(E_\nu)$ is
described as a superposition of
$L^2$-type five-body Gaussian basis functions 
written in all the Jacobi coordinates $c=1-5$ as shown in Fig.1.
By employing 15,000 five-body 
basis functions, {\it i.e.}, $\nu_{\rm max}=15,000$, 
the eigenfunction set $\{\Phi^{(\nu)}_{J^\pi M}(E_\nu)\}$ 
forms a nearly complete set in the finite interaction region.

The eigenfunctions $\Phi^{(\nu)}_{J^\pi M}(E_\nu)$ stand for
discretized continuum states of the five-body system, 
and are called ``pseudostates" in the scattering theory.
It is known that most of the pseudostates do not 
actually represent  resonance states 
but melt into non-resonant continuum states when the
scattering boundary condition is imposed to the total wave function. 

We actually perform the precise quark-model calculation 
including all the pseudostate terms in (2.2), 
and calculate the phase shift $\delta$ 
in the NK elastic scattering N+K$\to$N+K 
for $J^\pi=\frac{1}{2}^\pm, \frac{3}{2}^\pm$ channels, respectively.
For $J^\pi=\frac{1}{2}^\pm$, 
all the pseudostates located at $0 - 450$ MeV above the NK threshold 
melt into non-resonant continuum state when the NK scattering
boundary condition is imposed, which means the strong coupling
between those pseudostates and the NK scattering state.
We show in Fig.2 the calculated phase shift $\delta$, 
and find no resonance for $0 - 450$ MeV above the NK threshold, 
{\it i.e.,} $1.4-1.85 $ GeV in mass region around ${\rm \Theta}^+(1540)$. 
We find a sharp resonance for $J^\pi=\frac{1}{2}^-$ and a broad
one for $J^\pi=\frac{1}{2}^+$ around 1.9GeV in Fig.2, although 
their energies are too high to be identified with the ${\rm \Theta}^+(1540)$.
For the $J^\pi=\frac{3}{2}^\pm$ states, 
we find no resonance up to 500 MeV above the NK threshold.

\begin{figure}[h]
\begin{center}
\epsfig{file=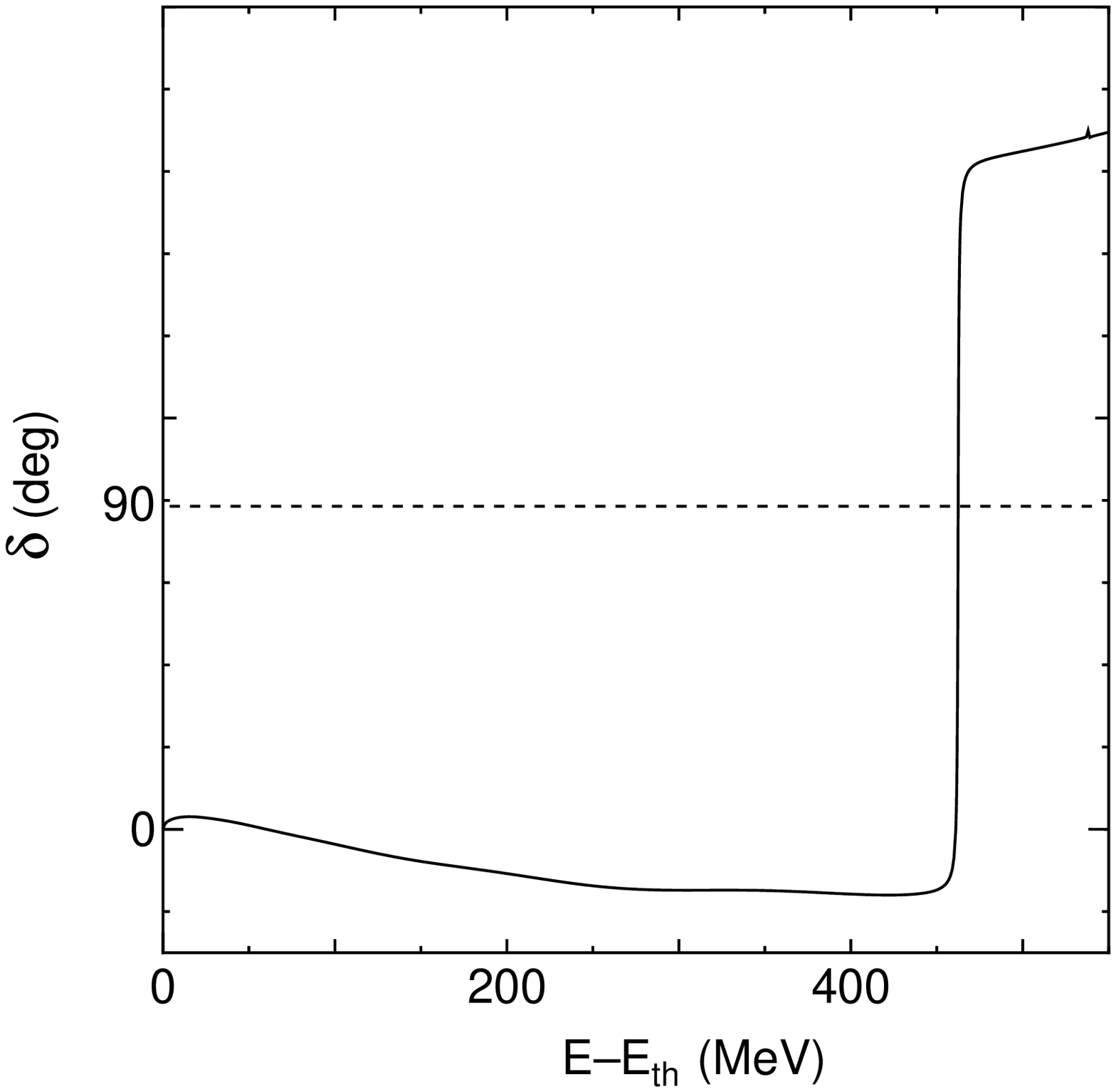,scale=0.28}
\hspace{0.5cm}
\epsfig{file=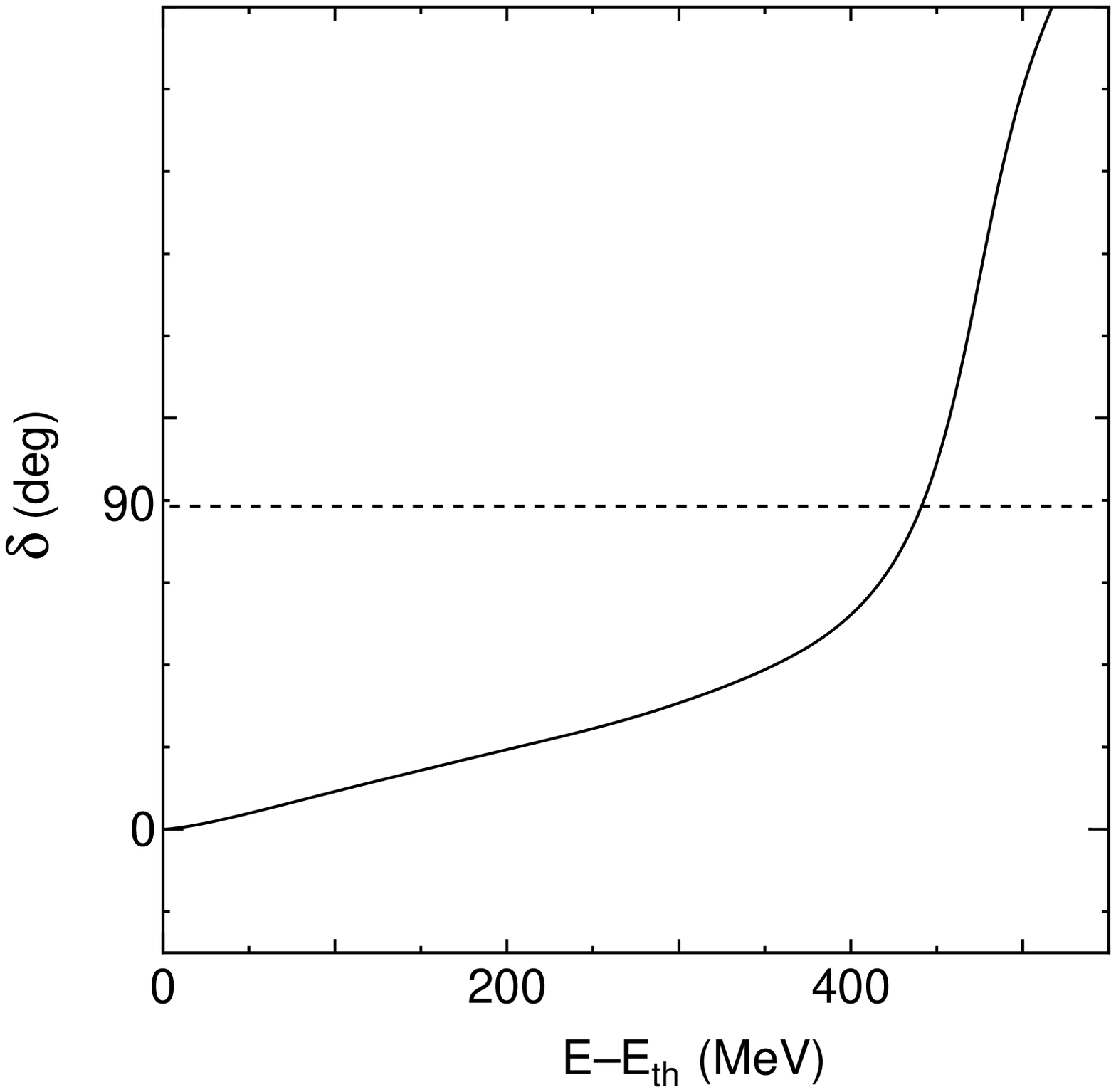,scale=0.28}
\end{center}
\caption{
The calculated phase shift $\delta$ for the five-quark system with 
$J^\pi=\frac{1}{2}^-$ (left) and $J^\pi=\frac{1}{2}^+$ (right) 
in the NK scattering (N+K $\to$ N+K) 
as the function of the energy measured from the NK threshold $E_{\rm th}=m_{\rm N}+m_{\rm K}$.
No resonance is seen around the energy of ${\rm \Theta}^+(1540)$.
Instead, there is a sharp resonance state with $J^\pi=\frac{1}{2}^-$ 
and the mass of about 1.9GeV.
}
\label{fig:phaseshifts}
\end{figure}
 
Note that these results on the absence of low-lying pentaquark resonances 
with $J^\pi=\frac{1}{2}^\pm, \frac{3}{2}^\pm$ 
are consistent with the lattice QCD results in Refs.9) and 10). 
As for the presence of $\frac{1}{2}^-$ penta-quark resonance around 1.9GeV, 
Ref.11) shows a consistent lattice QCD result 
indicating a $\frac{1}{2}^-$ penta-quark resonance around 1.8GeV.

To conclude, there is no five-quark resonance with $J^\pi=\frac{1}{2}^\pm, \frac{3}{2}^\pm$ below 1.85GeV.
Instead, the quark-model calculation predict a five-quark resonance state of $J^\pi=\frac{1}{2}^-$
with the mass of about 1.9GeV and the width of $\Gamma \simeq$ 2.68MeV.

\section{Four-body quark-model calculation for X$^0$(3872)}

With GEM, we analyze four-quark systems $c{\bar c}q{\bar q}$ for $I=0,1$  
in a quark model to investigate X$^0$(3872)\cite{X3872}, 
which may have the tetraquark structure as $c{\bar c}u{\bar u}$ \cite{Swanson}.
Note here that, if X$^0$(3872) is an $I=1$ state,
it is manifestly exotic and there should exist its isospin partners 
X$^\pm$ ($c{\bar c}u{\bar d}$ and $c{\bar c}d{\bar u}$) around 3.87GeV.
For the calculation of four-quark charmed systems, 
we adopt the quark-quark interaction of Ref.13), 
which effectively includes the exchange effect of Nambu-Goldstone bosons. 
With $m_c \simeq$ 1752MeV, $m_{u,d} \simeq$ 313MeV and $m_s \simeq$ 555MeV,
the quark model leads light hadron masses as 
$m_\rho\simeq$ 772.8MeV, $m_\omega\simeq$ 696.3MeV and $m_\pi\simeq$ 148.7MeV, 
and the calculated (experimental) masses of charmed mesons are 
$m_{\rm D}\simeq$ 1897.6(1867)MeV, $m_{\rm D^*}\simeq$ 2017.1(2008)MeV, 
$m_{{\rm J}/\psi}\simeq$ 3096.5(3097)MeV and $m_{\eta_c}\simeq$ 2989.1(2980)MeV. 

For the four-quark calculation, 
we employ all the 18 sets of the Jacobi coordinates of the four-body system.\cite{Hiyama2003} 
We show in Fig.4 the three important sets describing 
the $c{\bar u}$ and ${\bar c}u$ correlations ($c=1$), 
the $c{\bar c}$ and $u{\bar u}$ ones ($c=2$) and 
the $cu$ and ${\bar c}{\bar u}$ ones ($c=3$).
The four-body Gaussian basis functions are 
prepared in the Jacobi sets $c=1-18$ and
the four-body pseudostates
$\Phi^{(\nu)}_{J^\pi M}(E_\nu)$ are obtained
by diagonalizing the four-quark Hamiltonian
using  nearly 13,000 Gaussian basis functions.
For the phase shift calculation, we consider the scattering channels 
${\rm D}^0+{\rm \bar D}^{*0}$ and J/$\psi$ + $\omega$ for $I=0, J^\pi=1^+$,
and ${\rm D}^0+{\rm \bar D}^{*0}$ and J/$\psi$ + $\rho$ for $I=1, J^\pi=1^+$.

For the $I=0$ states, we find a very sharp resonance,
dominantly having the ${\rm D}^0 {\rm \bar D}^{*0}$ component, 
slightly below the ${\rm D}^0 \,{\rm \bar D}^{*0}$ threshold.
In this calculation, however, the J/$\psi$ $\omega$ threshold energy 
is  much lower than the experimental value by 86 MeV.
We find  that if the quark-quark interaction 
is artificially adjusted so as to reproduce 
the experimental value of the J/$\psi$ $\omega$ threshold, 
neither bound nor resonance state appears.
In fact, the result seems rather sensitive to the quark-quark interaction, 
and we need better interaction to obtain definite conclusions for $I=0 $ states.

For the $I=1$ states, we obtain several pseudostates near the
${\rm D}^0 {\rm \bar D}^{*0}$ threshold region, but all of them disappear
when the scattering boundary condition is switched on.
The resultant scattering phase shift by the full coupled-channel calculation
is given in Fig.3, which shows no resonance behavior.
We thus find no $c\bar cq\bar q$-type tetraquark resonance 
with $I=1$ in mass region of 3.87$-$4.0GeV.

\begin{figure}[ht]
\begin{center}
\epsfig{file=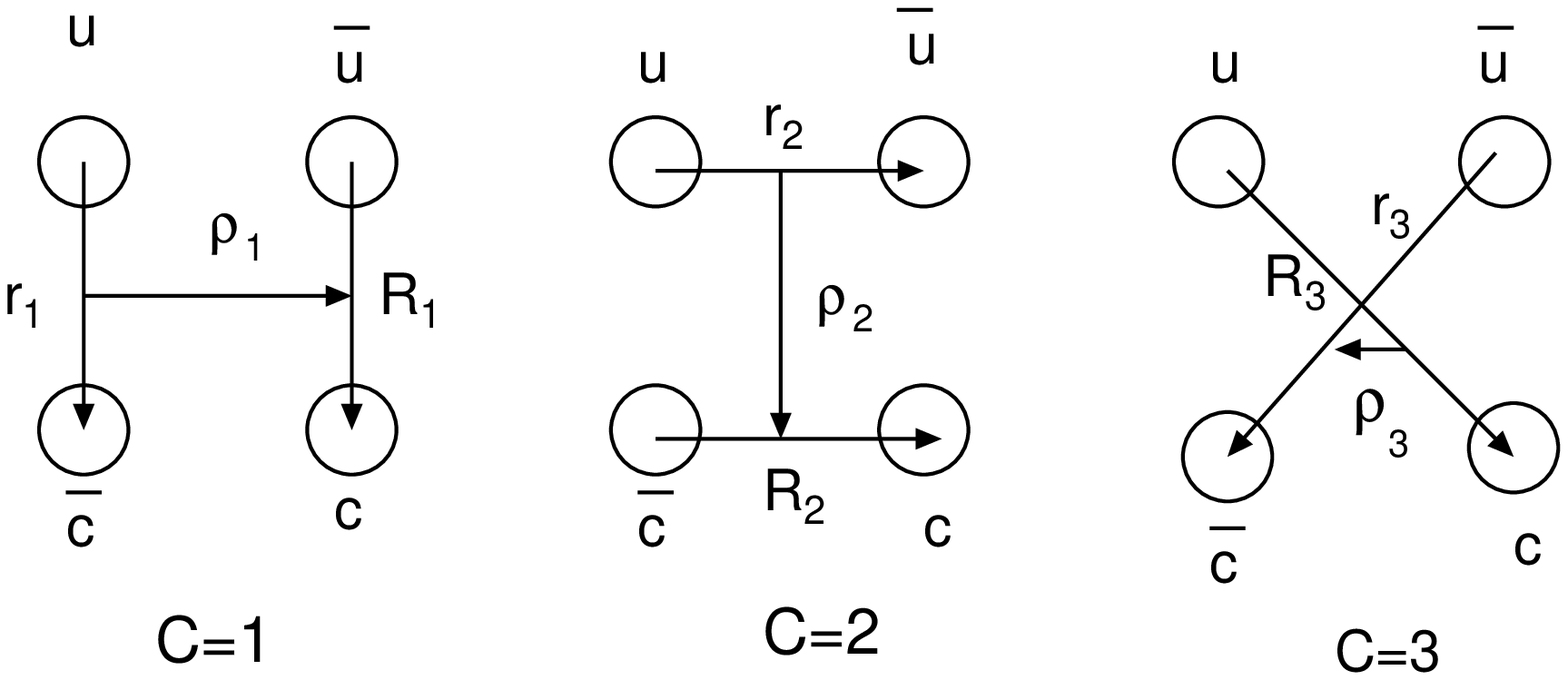,scale=0.43}
\hspace{0.5cm}
\epsfig{file=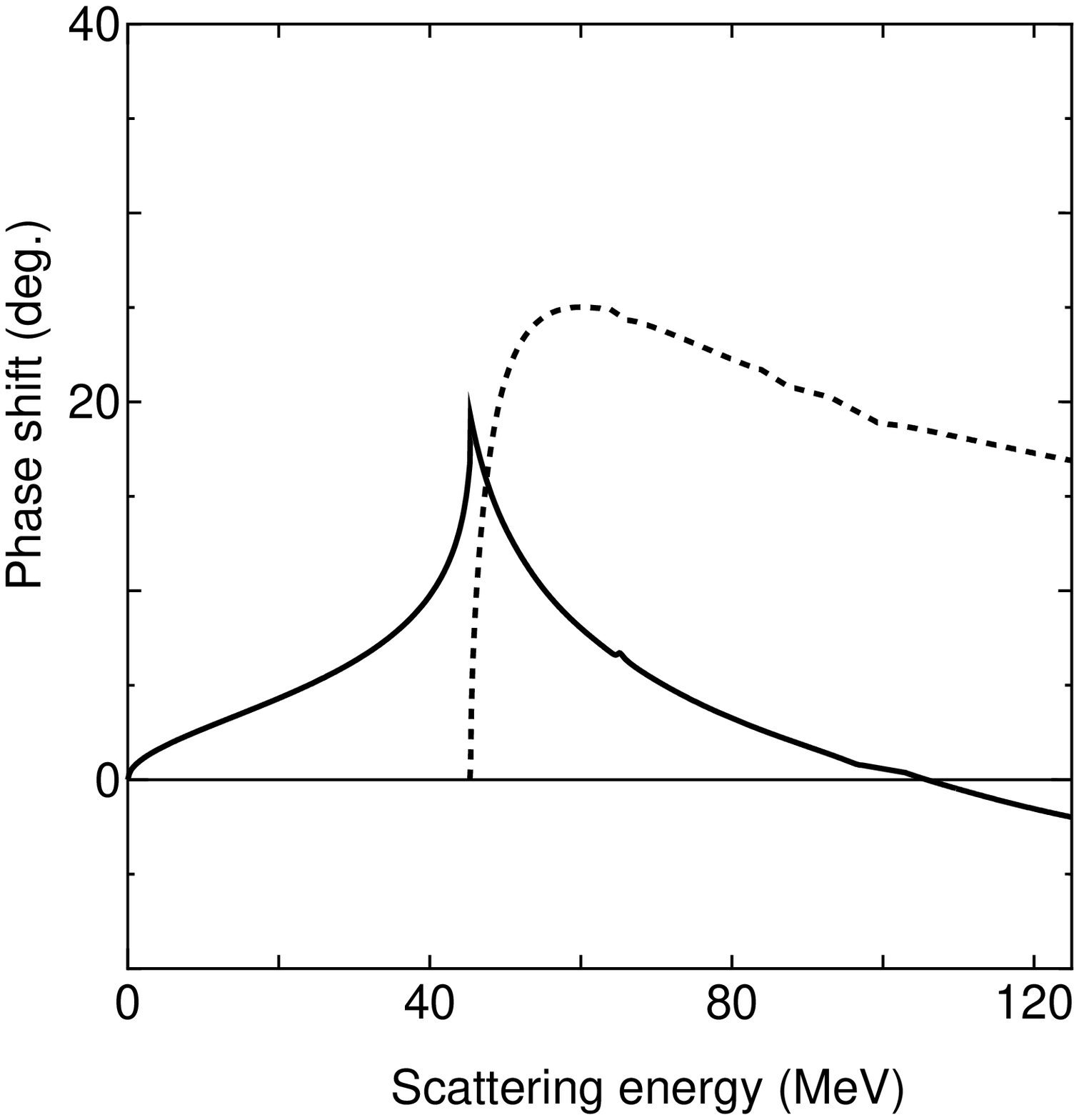,width=5.5cm,height=4.3cm}
\end{center}
\caption{(Left) Three important sets of the
Jacobi coordinates in the scattering calculation on
X$^0$(3872).  (Right) The phase shift in hadron-hadron scattering 
obtained by the four-body coupled-channel calculations 
for $c\bar c q\bar q$ four-quark systems with $I=1, J^\pi=1^+$: 
the solid curve for J/$\psi$ + $\rho \to$ J/$\psi$ + $\rho$ and 
the dotted curve for ${\rm D}^0+{\rm \bar D}^{*0} \to {\rm D}^0+{\rm \bar D}^{*0}$. 
The scattering energy is measured
from the J/$\psi$  $\rho$ threshold. 
The ${\rm D}^0{\rm \bar D}^{*0}$ threshold is at 45.3 MeV.
No resonance is seen although a threshold 
cusp appears due to the couple-channel effect. 
}
\label{fig:phaseshifts}
\end{figure}

In this way, the almost precise quark-model calculation 
using the Gaussian Expansion Method (GEM) is a powerful  tool 
to clarify the state properties of multi-quark systems, 
which is a theoretical search for exotic hadron resonances. 

\section*{Acknowledgements}
E.H. acknowledges Profs. A. Hosaka, H. Toki and M. Yahiro 
for useful discussions. 
The authors are thankful to the Yukawa 
Institute for Theoretical Physics at Kyoto University, 
where stimulating discussions were made about the present
work during the YKIS2006 on ``New Frontiers on QCD".

\end{document}